\documentclass{ws-ijmpe}

\begin{document}

\markboth{O.A.Grachov, B.Metzler, M.J.Murray et al.}{Measuring Photons and Neutrons at Zero Degrees in CMS}

\catchline{}{}{}{}{}

\title{Measuring Photons and Neutrons at Zero Degrees in CMS}
\author{O.A.Grachov, B.Metzler, M. Murray, J.Snyder, L.Stiles, J.Wood, V.Zhukova}
\address{University of Kansas, Lawrence, KS, 66045, USA}
\author{W.Beaumont, S.Ochesanu}
\address{Universiteit Antwerpen, Wilrijk, Belgium}
\author{P.Debbins, D.Ingram, E.Norbeck, Y.Onel}
\address{University of Iowa, Iowa City, IA, USA}
\author{E.Garcia}
\address{University of Illinois at Chicago, Chicago, IL, USA}
\maketitle

\begin{history}
\received{(received date)}
\revised{(revised date)}
\end{history}

\begin{abstract}

The CMS Zero Degree Calorimeters, ZDCs, will measure photons and neutrons emitted with
$|\eta|\geq 8.6$ from Pb+Pb, p+Pb and p+p collisions at $\sqrt{s_{NN}}=5.5$, $8.8$ and
$14$ TeV respectively. The calorimeter consists of an electromagnetic part segmented
in the horizontal direction and an hadronic part segmented into four units in depth.
In addition CMS will have access to data from a segmented shower maximum detector
being built for luminosity measurements. We will present detailed results from tests
beam measurements taken at the CERN SPS. These data will be used to extrapolate the
utility of the ZDCs to measure photons, and possibly $\pi_0$ s in p+p and
ultra-peripheral heavy-ion collisions. Data from the hadronic section can be used to
estimate the number of participants in heavy ion-collisions. In addition we will
discuss plans to use the detector to measure the reaction plane, thereby extending
the sensitivity of the central detectors in CMS.

\end{abstract}

\section{Introduction}
The Compact Muon Spectrometer (CMS) [1] consists of a tracking system,
electromagnetic and hadronic calorimeters and muon detectors. A solenoidal
magnet provides a 4 T magnetic field surrounding the tracking and calorimetric
systems. The tracking system covers the pseudorapidity region $|\eta|< 2.5$. The
combination of electromagnetic and hadronic calorimeters provides coverage of
the rapidity region $|\eta|<5$. The muon detector covers the region $|\eta|< 2.4$.
A set of two zero degree calorimeters (ZDCs) [2], with pseudorapidity coverage
of $|\eta|\geq 8.6$ for neutral particles, have been proposed to complement the existing
CMS detector, especially, for heavy ion studies. Each ZDC has two
independent parts: the electromagnetic (EM) and hadronic (HAD) sections.
Two identical ZDCs will be located between the two LHC beam pipes at ~140 m on each
side of the CMS interaction region at the detector slot of 1 m length, 96 mm width
and 607 mm height inside the neutral particle absorber TAN [3].

 In order to meet the goals of the CMS forward physics program the energy resolution of the  EM section must be on the level of $10\%$ for 50 GeV photons. During heavy ion running the 
combined (EM + HAD) calorimeter  should allow one to reconstruct the energy of 2.75 TeV
spectator neutrons with a resolution of $10-15\%$.

\section{Zero degree calorimeter}
Sampling calorimeters using tungsten and quartz fibers have been chosen
for the detection of the energy in the ZDCs. A significant advantage of this
technology is that the calorimeter will be very compact, extremely fast and radiation
hard. Quartz fibers were chosen as the active media of the ZDC calorimeters because of
their unique radiation hardness features and intrinsic speed of the Cherenkov effect.
The quartz-quartz fibers can withstand up to 30 GRad with only a few percent loss in
transperency in the wavelength range of 300-425 nm.\\ The HAD section consists of 24 layers
of 15.5 mm thick tungsten plates and 24 layers of 0.7 mm diameter quartz fibers $(6.5\lambda{o})$.
The tungsten plates are tilted by ${45}^0$.
The EM section is made of 33 layers of 2 mm thick tungsten plates and 33 layers of
0.7 mm diameter quartz fibers (19Xo). The tungsten plates are oriented vertically.
The fibers were laid in ribbons. The hadronic
section of each ZDC requires 24 fiber ribbons. After exiting the tungsten plates the fibers
from 6 individual ribbons are grouped together to form a read - out bundle. 
This bundle is compressed and glued  with epoxy into a tube. From there,
an optical air-core light guide will carry the light through radiation shielding to
the photomultiplier tube. The full hadronic section will consist of four identical  towers divided in the longitudinal direction. For the electromagnetic section, fibers from all 33 fiber ribbons
will be divided in the horizontal direction into five identical fiber bundles. These
5 bundles will form five horizontal towers, each fiber bundle will be mounted with a
0.5 mm air gap from the photocathode of a phototube. EM and HAD sections will be instrumented
with the same type of phototube. The phototubes are Hamamatsu R7525
phototubes with bi - alkali photocathode.

\begin{figure}[th]
\centerline{\psfig{file=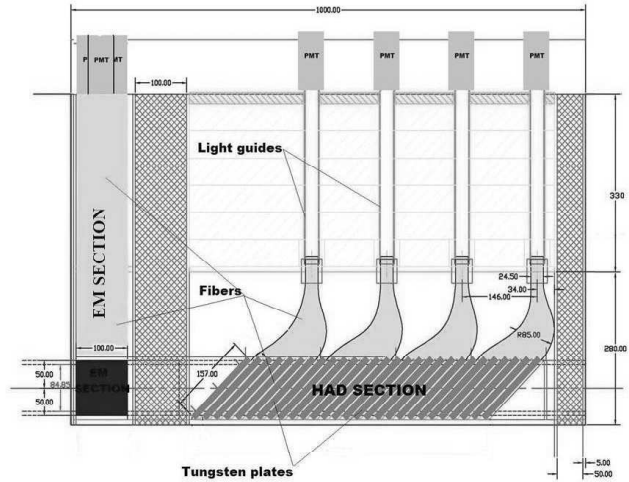,width=8cm}}
\vspace*{15pt}
\caption{The side view of the ZDC with EM section in front and
HAD section behind. A luminosity monitor will be mounted in the 10 cm space
between the ZDC's calorimetric sections}
\end{figure}

Figure 1 shows a side view of the ZDC with the EM section in front and 
the HAD section behind. The configuration includes 9 mm Cu plates in the front and back of each section.
For the TAN's final detector configuration a luminosity monitor[4] will be mounted in the 10 cm space
between the ZDC's calorimetric sections.

\section{Experimental set-up}
The main goal of the test beam measurements was to study the performance of the
zero degree calorimeter. Secondary goals included tests of the full electronic chain to be used in 
CMS and real data data transfer to Fermi National Accelerator Lab for online monitoring.  
Measurements were carried out in the SPS H2 beam at CERN.
A 400 GeV/c proton beam was extracted
from the SPS accelerator and steered onto the primary target. The intensity  of this
primary beam was $~ 5 \times 10^{12}$ protons per burst. Downstream from the primary
target, a secondary hadron or electron (positron) beam were made with momenta from
10 to 350 GeV/c. 
The beam  then passed through
a gas threshold Cherenkov counter; two multi - wire proportional chambers MWPCs, of size $({10}\times{10}) {cm}^2$;  
 four 1cm thick scintilators with sizes 
 $({14}\times{14}) {cm}^2$
, $({4}\times{4}) {cm}^2$, $({2}\times{2}) {cm}^2$, and  $({14}\times{14}) {cm}^2$ and finally
two more wire chambers (WC-D and WC-E),  
before hitting the calorimeter.
The calorimeter sat on table that could be moved horizontal and vertical directions under remote control.  This made it possible to direct the beam anywhere in the calorimeter
front face.\\

The trigger system required a coincidence from the scintillaters and the appropriate signal in the cerenkov.    
The normal beamspot condition exposed an calorimeter area of about $({2}\times{2}) {cm}^2$.
The MWPCs were 
used to extrapolate the particle trajectory to the calorimeter front face.

The high-voltage for the phototubes was provided by a commercial unit LeCroy4032 \cite{HfElectronics}. Each tube was pre-calibrated. Based on pr-ecalibration data and simple calculations we set the same gain of $10^{5}$ for each PMT.
The electrical signals from the phototubes were transmitted through a 204 m long coaxial cables type
C-50-11-1 to charge integrater and encoder (QIE) for digitization and buffering. All of the digital electronics ran at 40 MHz, the clock speed that will be used at the LHC. 

\subsection{Response to Electrons}
The characteristics of the EM section were measured with 20, 50, 100 Gev positrons. The
momentum spread of the beam was 0.1-0.2\%. To equalize each tower the center of each tower
was irradiated by 50 GeV positron beams. The peak position obtained from a
Gaussian fit of the amplitude distribution for each tower was used to determine the
calibration coefficients.\\
\begin{figure}[th]
\centerline{\psfig{file=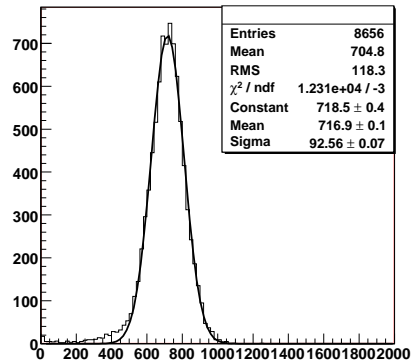,width=6cm}}
\vspace*{8pt}
\caption{The sum of the gain corrected signals in the central tower irradiated by
50 Gev/c positrons and the two neighbouring towers.}
\end{figure}
Figure 2 shows the sum of the response of the central tower and its two neighbours to  
50 Gev/c positrons. The energy resolution for the different positron energies measurements was obtained by Gaussian fits and can be parametrized as,
\begin{equation}
\left(\frac{\sigma}{E}\right)^2=\left(\frac{70\%}{\sqrt{E}}\right)^2 + \left({8\%}\right)^2
\end{equation}
The linearity of the EM section response is defined as the ratio of the fitted mean energy to
the nominal beam energy. The relative deviation from a straight line as a function of beam
momentum have been measured and the calorimeter is found to be linear over a range of from 20 GeV
to 100 GeV to within 2\%. The GEANT4 simulations reproduces well the EM section test beam data.\\

\subsection{Response to pions}
Positive pions with energies of 150 GeV and 300 GeV were used to measure the response of the EM + HAD combined system.
The total depth of combined system is ~7.5 hadronic interaction lengths $(\lambda)$.\\ The total energy:
\begin{equation}
{E}_{TOT}=\alpha{E}_{EM}+{E}_{HAD}
\end{equation}
 is defined as the sum of energy in EM section:
\begin{equation}
{E}_{EM}={E}_{EM1}+{E}_{EM2}+{E}_{EM3}+{E}_{EM4}+{E}_{EM5}
\end{equation}
and energy in HAD section:
\begin{equation}
{E}_{HAD}={E}_{HAD1}+{E}_{HAD2}+{E}_{HAD3}+{E}_{HAD4},
\end{equation}
 when the central tower (EM3) was irradiated by beam.
\begin{figure}[th]
\centerline{\psfig{file=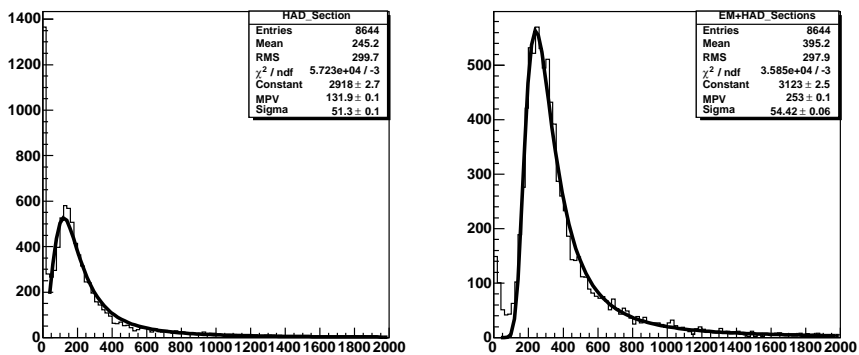,width=12cm}}
\vspace*{8pt}
\caption{HAD section and combined EM+HAD system response to 300 GeV pions}
\end{figure}
The energy dependent intercalibration parameter between the EM and HAD section $(\alpha)$ is determined by minimizing the energy resolution of 300 GeV pions. In our case, $\alpha$ is equal 1 at 300GeV. The analysis did not included any corrections for dead material, leakage and the non-compensation nature of calorimter's sections.
Figure 3 shows an example of HAD section and combined EM+HAD system response to 300 GeV pions.
The energy resolution was obtained by a Landau fit and can be parametrized as,
\begin{eqnarray}
\frac{\sigma}{E}=\frac{138\%}{\sqrt{E}} + {13\%}
\end{eqnarray}

\section{Flow detector up-grade}
The study of directed flow has been one of the main tools used to
study the strongly interacting matter created at RHIC. It has been
suggested that if a quark gluon plasma is created the directed flow,
as measured by $v_1$, should oscillate, or wiggle, as one moves from
central to forward rapidities $[5,6]$. At forward rapidities the plasma
may expand in the direction
opposite the reaction plane. The STAR collaboration has searched for
this effect at
$\sqrt{s_{NN}}$ = 62.4 GeV/c using their shower maximum detector,
imbedded in the Zero Degree Calorimeters to reconstruct the reaction
plane $[7]$. So far the ``wiggle" has not been seen in data,
possibly because we have not yet reached the softest point of the
equation of state.
The CMS experiment has the largest (psuedo)-rapdity coverage of any
experiment at the LHC. The complete azimuthal coverage and fine
granularity of the detectors make it ideal to study the rapidity
dependence of flow.
We propose to augment these capabilities by adding a flow detector to
the
CMS Zero Degree Calorimeters. These devices will consist of two 8cm
by 8cm  hodoscopes inserted just behind the electromagnetic sections
of the left and right ZDCs. We are currently investigating a variety
of technologies to  address the technical challanges involved in this
project.

\section{Summary}
The ZDC has sufficient energy resolution and linearity to meet our physics goals. The response also agrees well with GEANT4 simulations. The electromagnetic section has a non-linearity $\le 2\%$ from 
20 to 100GeV. EM section, as an independent detector, significantly improves the energy resolution for hadrons with energy between 150 and 300 GeV. The combined resolution of electromagnetic 
 and hadronic sections is approximately twice that of the hadronic section alone.\\

\section{Acknowledgements}
The results presented in this paper are due to the collaborative work of many people from the CMS Collaboration, and it is our pleasure to acknowledge their important contribution. The detector's studies were only possible due to excellent machine operation by the SPS crew at CERN.

\end{document}